\documentclass[10pt,conference,letterpaper]{IEEEtran}

\IEEEoverridecommandlockouts

\usepackage{amsmath,amssymb,amsfonts,mathtools,xparse}
\usepackage{textcomp}
\usepackage{xcolor}
\usepackage{cite,epsfig,graphicx}

\usepackage{multicol}

\usepackage{algorithm}
\usepackage{algcompatible}
\usepackage[noend]{algpseudocode}

\usepackage{tabularx}

\graphicspath{{image/}}

\usepackage{scalerel}
\newcommand {\inlinefig}[1] {\scalerel*{\includegraphics{#1}}{|}}

\algnewcommand\Not{\textbf{not}}
\algnewcommand\Or{\textbf{or}}
\makeatletter
\algnewcommand{\LineComment}[1]{\Statex \hskip\ALG@thistlm \(\triangleright\) #1}
\makeatother
\algdef{SE}[DOWHILE]{Do}{doWhile}{\algorithmicdo}[1]{\algorithmicwhile\ #1}%

\usepackage{url}

\makeatletter
\def\endthebibliography{%
	\def\@noitemerr{\@latex@warning{Empty `thebibliography' environment}}%
	\endlist
}
\def\BState{\State\hskip-\ALG@thistlm}
\makeatother

\begin{document}
\title{Analysis of Attacker Behavior in Compromised Hosts During Command and Control\thanks{This research is supported by the National Science Foundation (NSF), USA, Award \#1739032.}}

\author{
	\IEEEauthorblockN{
		Farhan Sadique\IEEEauthorrefmark{1},
		Shamik Sengupta\IEEEauthorrefmark{2}
	}
	\IEEEauthorblockA{
		\textit{Dept. of Computer Science and Engineering}  \\
		\textit{University of Nevada, Reno}, NV, USA \\
		fsadique@nevada.unr.edu\IEEEauthorrefmark{1},
		ssengupta@unr.edu\IEEEauthorrefmark{2}
	}
}

\IEEEoverridecommandlockouts

\maketitle
\IEEEpubidadjcol

\begin{abstract}

Traditional reactive approach of blacklisting botnets fails to adapt to the rapidly evolving landscape of cyberattacks. An automated and proactive approach to detect and block botnet hosts will immensely benefit the industry. Behavioral analysis of botnet is shown to be effective against a wide variety of attack types. Current works, however, focus solely on analyzing network traffic from and to the bots. In this work we take a different approach of analyzing the chain of commands input by attackers in a compromised host. We have deployed several honeypots to simulate Linux shells and allowed attackers access to the shells to collect a large dataset of commands. We have further developed an automated mechanism to analyze these data. For the automation we have developed a system called CYbersecurity information Exchange with Privacy (CYBEX-P). Finally, we have done a sequential analysis on the dataset to show that we can successfully predict attacker behavior from the shell commands without analyzing network traffic like previous works.

\end{abstract}


\section{Introduction}\label{sec:intro}


A key component in cyberattacks is a bot -- a compromised host or a botnet \cite{puri2003bots, feily2009survey} -- a set of such hosts. A host becomes a bot when an attacker gains access to its shell. To do that the attacker can plant a malware (e.g. mirai \cite{antonakakis2017understanding}, Torpig \cite{stone2009your}, Conficker \cite{shin2010conficker} etc.) or directly perform brute-force attack to crack the password. Attackers use botnet to distribute malware, perform DDoS attacks, host phishing websites, perform brute-force attacks etc. Bots make up a large portion of the cybersecurity market. So it is desirable to detect and block a bot in any network.

\subsection{Motivation \& Challenges}

Currently the most popular defense against botnet is manually blacklisting their IP addresses. However, numerous hosts are compromised every day. At the same time, many bots become benign as the owner regains control of the host. So, it is impossible to list all their IPs. Moreover, blacklisting is a reactive approach. An IP shows up in a blacklist only after it has done some harm.

As a result, the industry would greatly benefit from a proactive defense mechanism against botnets. An intelligent system should detect a zero-day bot from its behavior not the IP. If a bot is detected in an early phase of the kill chain, it cannot do any harm to anybody else.

There is extensive research on behavioral analysis of botnet. Intrusion detection systems (IDS) \cite{liao2013intrusion} use network signatures to detect bots. While they work really well for known patterns they cannot adapt to the new attacks. It also takes a long time to detect an attack pattern, analyze it and create its signature before adding it to an IDS.

Another popular approach is using anomalies in network traffic \cite{karasaridis2007wide, binkley2006algorithm, gu2008botsniffer} to detect bots. Some works take it further to detect anomalies in DNS traffic \cite{choi2007botnet, villamarin2008identifying, dagon2005botnet, schonewille2006domain}. However, network traffic data is cumbersome to work with and often encrypted.

\subsection{Contribution}

In this work, we take a novel approach of analyzing the chain of commands input by attackers in a compromised hosts, rather than using network traffic data. A safe but effective approach of documenting commands input in bots is to use honeypots \cite{provos2004virtual}.

We have allowed attackers access to a simulated shell of the Cowrie \cite{oosterhof2016cowrie} honeypot. After gaining access to the honeypot's shell, attackers input several commands. Attackers try to gain information about the system or gain access to a privileged shell. After trying out several commands the attacker logs out of the shell. We call this sequence of commands input during one particular login session a command chain.

We then create a model from these command chains based on their frequency occurrence. Finally, we try to predict the next command to be input by the attacker from our dataset. We have achieved an accuracy of 94-99\% in predicting the next command input by the attacker. This shows that attacker behavior is predictable using command chains only. To the best of our knowledge, no previous work has looked into this behavior of botnets and our approach is a novel one.


%

\section{Related Work}\label{sec:rel}

Extensive work was done on botnet behavior. Some of them used dataset similar to ours. Shrivastava et al. \cite{shrivastava2019attack} captured different attacks on IoT devices using the Cowrie \cite{oosterhof2016cowrie} honeypot. They employed various machine learning algorithms, including Naive Bayes, Random Forest and Support Vector Machine (SVM) to classify these attack. Surnin et al. studied different techniques  for SSH detection and proposed a methodology for probabilistic estimation of honeypot detection. Dowling et al. \cite{dowling2018using} used reinforcement learning to achieve a similar goal of concealing honeypot functionality.

Several works studied the Mirai \cite{antonakakis2017understanding} botnet using Cowrie honeypot. Kambourakis et al. \cite{kambourakis2017mirai} provided details on how the Mirai botnet  malware spreads, and discussed defensive  strategies. Said et al.\cite{said2018detection} discussed techniques to classify binary samples as Mirai based on their syntactic and behavioral properties. Lingenfelter et al. \cite{lingenfelter2020analyzing} analyzed variation among IoT Botnets using medium interaction honeypots. However, none of these works used our approach of analyzing the shell commands during the command and control stage. Our work differs by proposing an automated methodology to analyze their behavior by analyzing the shell commands input during the command and control stage.

Several previous works modeled the attacker behavior in IoT network. Deshmukh et al. \cite{deshmukh2019attacker}  proposes Fusion Hidden Markov Model (FHMM) for modeling attacker behavior. FHMM is more noise resistant and provides faster performance than  Deep  Recurrent  Neural  Network  (DeepRNN) with comparative accuracy.

Finally, Rade et al. \cite{rade2018temporal} modeled honeypot data using semisupervised Markov Chains and Hidden Markov Models (HMM). They also explored Long Short-Term Memory (LSTM) for attack sequence modeling. They concluded that LSTM provides better accuracy than HMM. As shown in section \ref{sec:res}, we have used normalized Levenshtein distance \cite{yujian2007normalized} to get a faster performance and a better accuracy. We also propose an automation framework to better capture the evolving nature of the threat landscape.

\section{System Architecture}\label{sec:sysarch}

\begin{figure}
	\includegraphics[width=\linewidth]{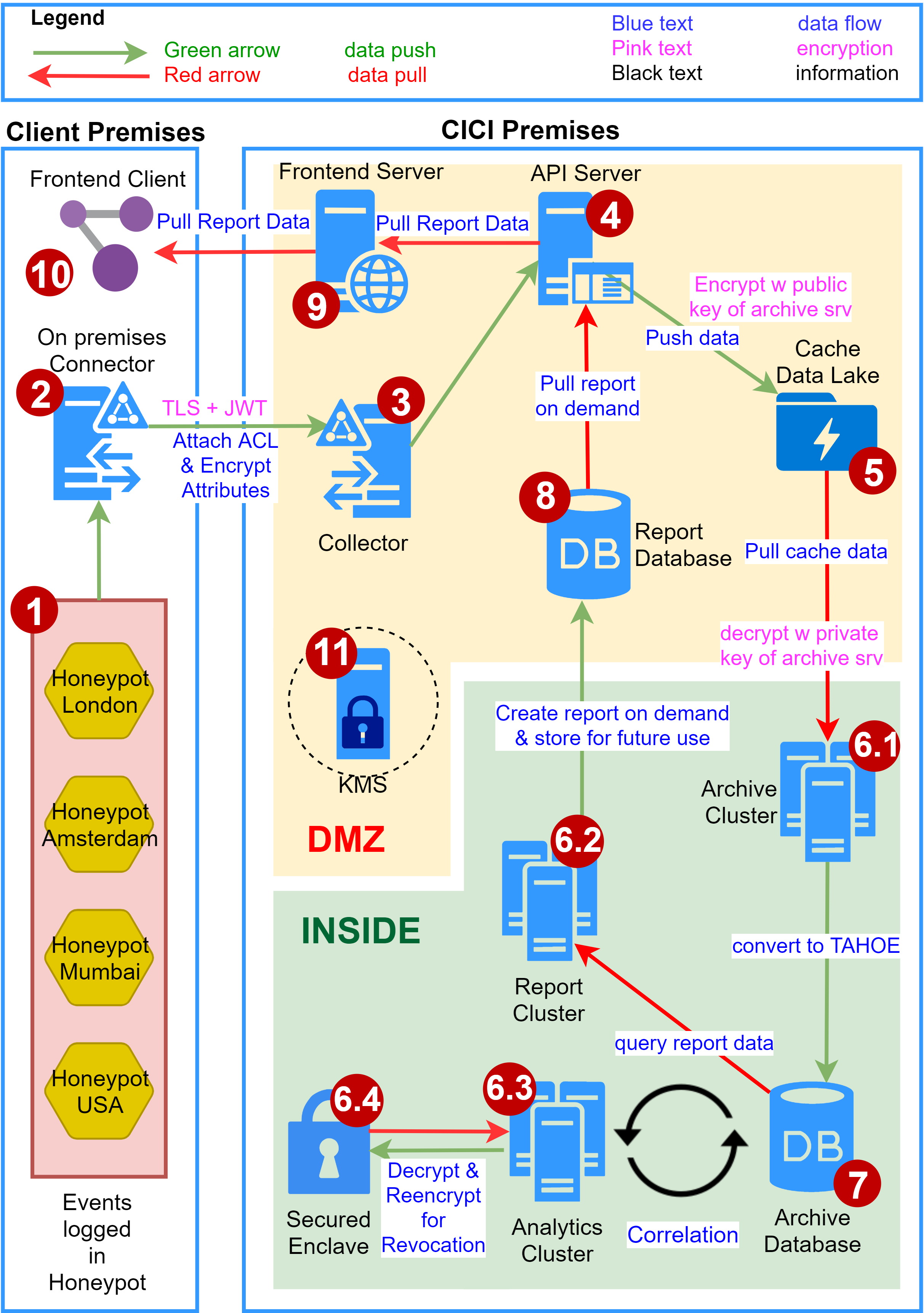}
	\centering
	\caption{System architecture of CYBEX-P along with the Data Flow.}
	\label{fig:sysarch}
\end{figure}

In this work, we have developed an automated framework to analyze and fingerprint attacker behavior in any compromised host. Our system uses CYbersecurity information Exchange with Privacy (CYBEX-P) \cite{sadique2019system} infrastructure as a service (IaaS). CYBEX-P is a cloud based platform for organizations to share heterogeneous cyberthreat data. CYBEX-P accepts all kinds of human or machine generated data including firewall logs, emails, malware signatures etc. Our system has $7$ modules -- (1) Honeypots, (2) Frontend, (3) Input, (4) API, (5) Archive, (6) Analytics, and (7) Report. These modules share various components as shown in Fig. \ref{fig:sysarch}.

\subsubsection{Honeypots}

We have setup $4$ instances of the Cowrie honeypot all around the world. The locations are -- Amsterdam, London, Mumbai and San Jose. All of them login SSH login attempts and corresponding commands input upon successful login. We have a diverse choice of four honeypots in four locations for better analysis and correlation.

\subsubsection{Frontend Module}\label{ss:fend}

The frontend module (\inlinefig{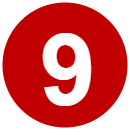}, \inlinefig{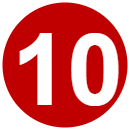} in Fig. \ref{fig:sysarch}) is a webapp for users to interact with CYBEX-P. This module allows users -- (1) to register with and login to CYBEX-P, (2) to configure the data sources, (3) to view the data, (4) to generate reports, and (5) to visualize the data.

\subsubsection{Input module}\label{subsec:input}

The input module (\inlinefig{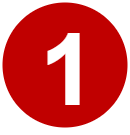}, \inlinefig{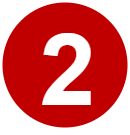}, \inlinefig{3}, \inlinefig{4}, \inlinefig{10} in Fig. \ref{fig:sysarch}) handles all the data incoming to CYBEX-P. Machine data is automatically sent via a connector (\inlinefig{2}) to the collector (\scalerel*{\includegraphics{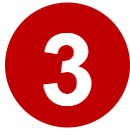}}{|}) using real time websockets. Afterwards, the collector posts the raw data to our API (\scalerel*{\includegraphics{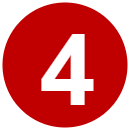}}{|}) endpoint. To ensure privacy, it uses the transport layer security (TLS) protocol \cite{dierks2008transport} during collection and posting.

\subsubsection{API module}\label{subsec:api}

The API module (\inlinefig{4}, \inlinefig{5} in Fig. \ref{fig:sysarch}) consists of the API server (\inlinefig{4}) and the cache data lake (\inlinefig{5}). It acts as the gateway for all data into and out of CYBEX-P. It serves two primary purposes:

\begin{enumerate}
	\item The input module (subsection \ref{subsec:input}) puts raw data into the system using the API.
	\item The report module (subsection \ref{subsec:report}) sends reports back to users using the API.
\end{enumerate}

\subsubsection{Archive module}\label{subsec:archive}

The archive module (\inlinefig{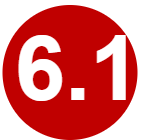}, \inlinefig{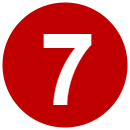} in Fig. \ref{fig:sysarch}) resides in the archive cluster and consists primarily of a set of parsing scripts. As mentioned earlier, the cache data lake (\inlinefig{5}) is encrypted with the public key of the archive server (\inlinefig{6_1}). The archive server -- (1) gets the encrypted data from the cache data lake (2) decrypts the data using own private key (3) parses the data into TAHOE, and (4) stores the data in the archive DB (\inlinefig{7}).

\subsubsection{Analytics module}\label{subsec:analytics}

The analytics module (\inlinefig{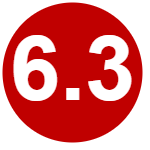}, \inlinefig{7} in Fig. \ref{fig:sysarch}) works on the archived data to transform, enrich, analyze or correlate them. It has various sub modules, some of which described here.

\paragraph{Filter sub-module}

An analytics filter parses a specific event from raw user data. Multiple filters can act on the same raw data and vice-versa.  For example, one filter can extract a \textit{file download event} from a piece of data while another filter can extract a \textit{DNS query event} from the same data.

\paragraph{Sequential Analysis sub-module}

This is a specialized sub-module that performs sequential analysis of the data based on the timestamp. It also correlates events in a session. A session is the time duration when one user is logged in.

\subsubsection{Report Module}\label{subsec:report}


Users use the report module (\inlinefig{4}, \inlinefig{5}, \inlinefig{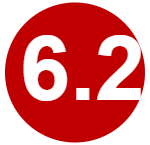}, \inlinefig{7}, \inlinefig{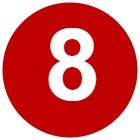}, \inlinefig{9}, \inlinefig{10} in Fig. \ref{fig:sysarch}) to generate and view reports. They request reports via the frontend client (\inlinefig{10},\inlinefig{9}). The API (\inlinefig{4}) stores the requests in the cache data lake (\inlinefig{5}). The report server (\inlinefig{6_2}) handles those requests by getting relevant data from the archive DB (\inlinefig{7}) and aggregating them into reports. It then stores the reports in the report DB (\inlinefig{8}). Users can access the reports on demand.

\section{Dataset}\label{sec:dataset}

\subsection{Data Source -- Cowrie Honeypot}

Cowrie \cite{oosterhof2016cowrie} is a medium to high interaction SSH and Telnet honeypot designed to log brute force attacks and the shell interaction performed by the attacker. In medium interaction mode (shell) it emulates a UNIX system in Python, in high interaction mode (proxy) it functions as an SSH and telnet proxy to observe attacker behavior to another system. In this work we use Cowrie in the medium interaction mode.

In our setup, cowrie only allows SSH logins into our honeypot. An attacker can login to the system using any username and password combination. Cowrie logs all the interactions including the source IP address of the attacker, the SSH parameters, the downloaded files and the commands input while the attacker is logged in.

\subsection{Cowrie Data as Events}

\begin{figure}[!h]
    \includegraphics[width=\linewidth]{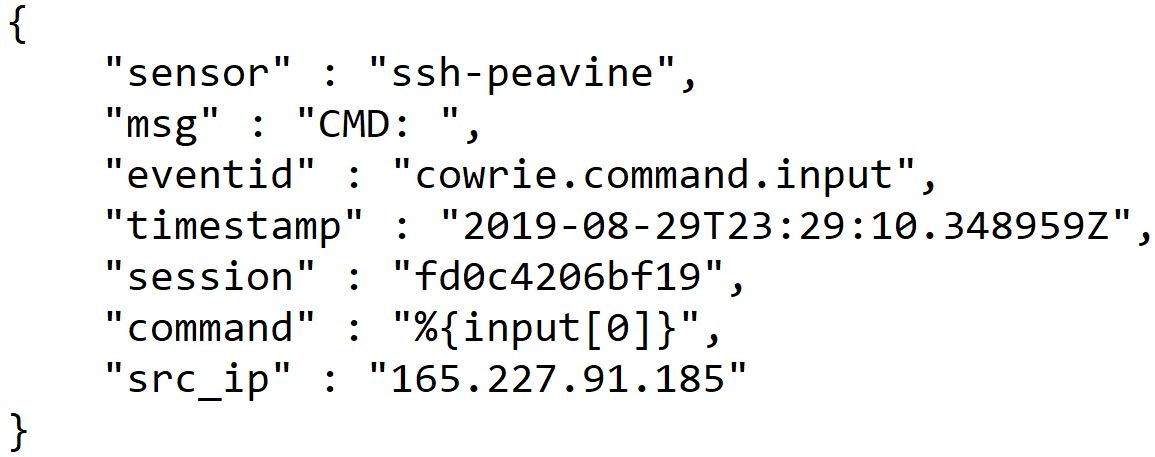}
    \centering
    \caption{Attributes of a Cowrie Event.}
    \label{fig:cowrie}
\end{figure}

Cowrie structures collected data into events. Fig. \ref{fig:cowrie} shows the attributes of a Cowrie event. Here

\begin{itemize}
    \item \textit{eventid:} Describes the type of the data, in the Fig. \ref{fig:cowrie}, \texttt{cowrie.command.input} means this event contains a command input by the attacker in the shell.
    \item \textit{timestamp:} The time when the event was recorded by the honeypot.
    \item \textit{msg:} The log message generated by Cowrie.
    \item \textit{src\_ip:} Source IP of the attacker.
    \item \textit{session:} Cowrie maintains a \texttt{session} id to group events logged during one login session by an attacker.
    \item \textit{sensor:} Host-name of the honeypot server.
    \item \textit{command:} This field is present only for eventid \texttt{cowrie.command.input} and contains the exact command input by the attacker.
\end{itemize}

Cowrie generates about $20$ different `eventid's. Many of these events are related to the SSH session, key exchange and data logging and do not carry valuable data. For this work, we are primarily interested in the `eventid' \texttt{cowrie.command.input}, which stores the exact command input by an attacker into the shell.

\subsection{Data Statistics}

For this work we have collected more than $18$ million Cowrie events, spanning about $700$ thousand sessions, from 27 August 2019 to 8 September 2019. Although we have $4$ honeypot installations our initial analysis shows that all of them faced the same type of attack. The datasets in all the honeypots were dominated by the Mirai \cite{kambourakis2017mirai} botnet. This is because Cowrie honeypot emulates an IoT device and during $2019$ Mirai was dominating IoT devices.

We have grouped all the events from all the honeypots together in one dataset because they were generated by the same malware (Mirai). Our initial analysis showed that there was no qualitative difference between the data generated from the different honeypots. So we have not done any comparative analysis in this work, rather analyzed all the data together.

\section{Analysis Methodology}\label{sec:meth}

As mentioned in section \ref{sec:sysarch}, We have used CYBEX-P \cite{sadique2019system} to automate the entire procedure of data collection to data analysis for this work. In other words we have used CYBEX-P infrastructure as a service (IaaS) here. This work is closely coupled with the development of CYBEX-P.

Along with CYBEX-P, we have further developed TAHOE a graph-based cyberthreat language (CTL). TAHOE offers several advantages over traditional CTLs. Firstly, TAHOE can store all types of structured data. Secondly, queries in TAHOE format are faster than in other CTLs. Finally, TAHOE scales well for analyzing data - a major limitation of other CTLs. Furthermore, it structures threat data as JavaScript Object Notation (JSON) which is very versatile compared to SQL.

\subsubsection{Data Generation}

Each Cowrie honeypot (\inlinefig{2} in Fig. \ref{fig:sysarch}) simulates a generic IoT device. They generate data in the format of Fig. \ref{fig:cowrie}. The honeypots log these data pieces into a file in respective server. We call each such log message a raw document.

\subsubsection{Data Input}

Each of our honeypot installations have a connector agent (\inlinefig{2} in Fig. \ref{fig:sysarch}). The connector is basically a script that reads the raw data from log files and sends them to the CYBEX-P collector (\inlinefig{3}) via a websocket. The data in transport are encrypted via TLS.

\subsubsection{Data Collection}\label{ss:datacoll}

The collector then posts the data to the API (\inlinefig{4}). The API encrypts the data with the public key of the archive cluster (\inlinefig{6_1}) and stores the encrypted data in the cache data lake (\inlinefig{5}).

\subsubsection{Data Archiving}

\begin{figure}
    \includegraphics[width=\linewidth]{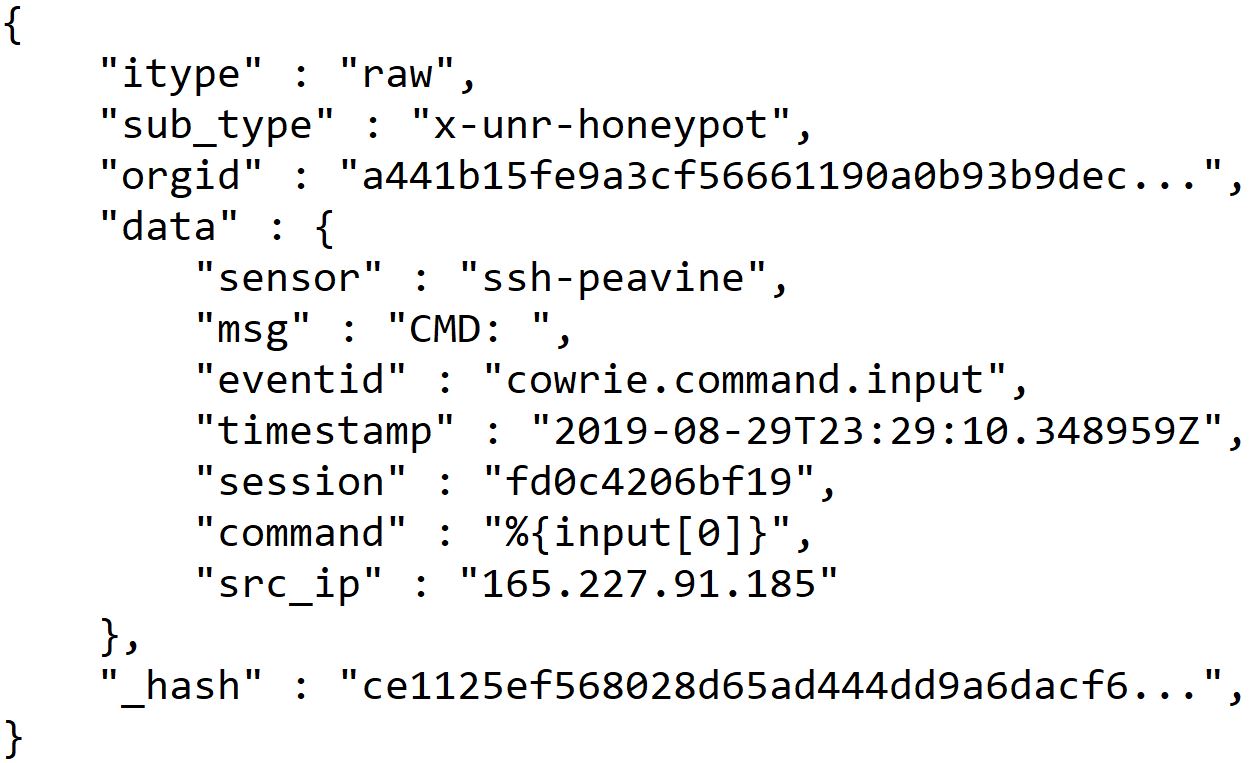}
    \centering
    \caption{A TAHOE \texttt{raw} document.}
    \label{fig:raw}
\end{figure}

The archive cluster (\inlinefig{6_1}), then pulls the data from the cache data lake (\scalerel*{\includegraphics{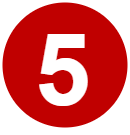}}{|}), decrypts the data using its private key, converts the cowrie events into TAHOE \texttt{raw} format and stores them in the archive database (\inlinefig{7}). TAHOE \texttt{raw} basically puts a wrapper around the Cowrie event. Fig. \ref{fig:raw} shows the structure of a TAHOE \texttt{raw} document.

\subsubsection{Data Analytics}

\begin{figure}
    \includegraphics[width=\linewidth]{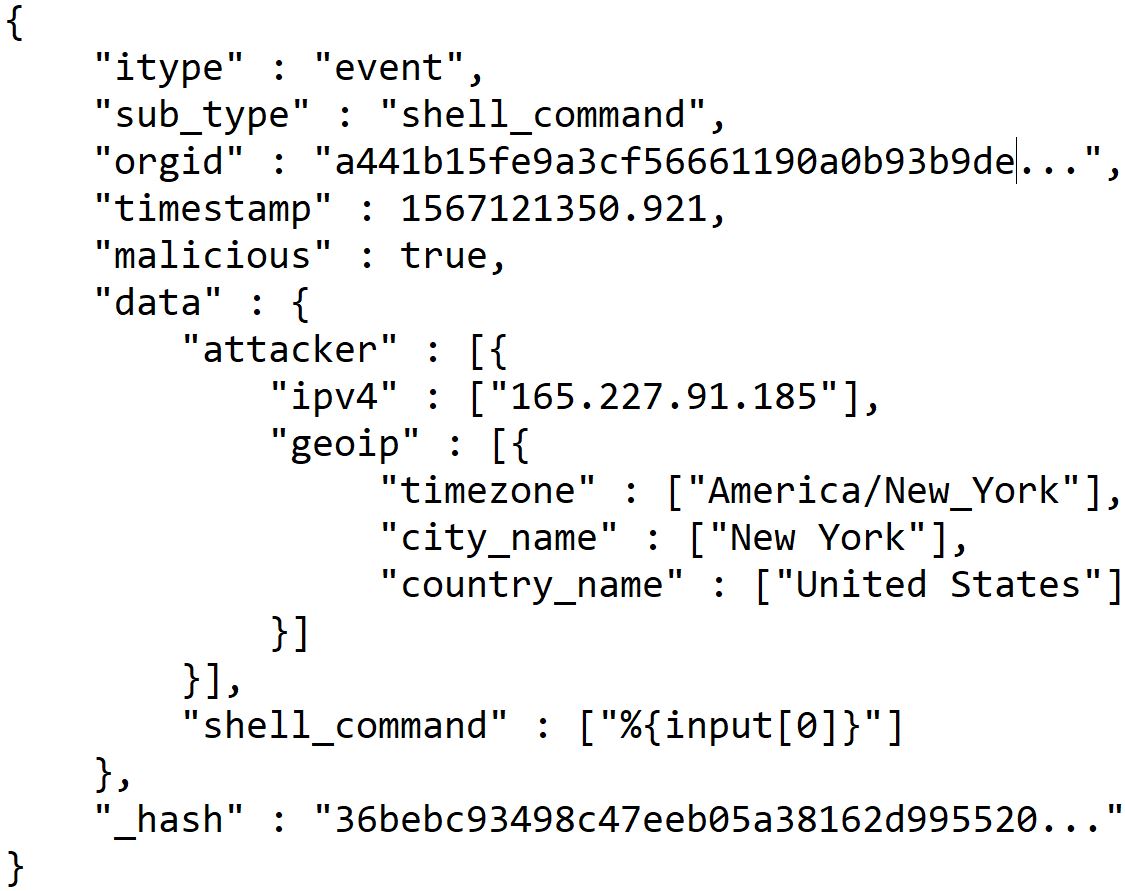}
    \centering
    \caption{A TAHOE \texttt{event}.}
    \label{fig:event}
\end{figure}

The analytics cluster (\inlinefig{6_3}) parses the TAHOE \texttt{raw} data into TAHOE \texttt{events} and \texttt{sessions}. It reads the \texttt{raw} data from the archive database (\inlinefig{7}), processes the data and writes the results back in the archive database. Fig. \ref{fig:event} shows structure of a TAHOE \texttt{event}.

%

\subsubsection{Command Sequence}\label{ss:seq}

\begin{figure}[h]
    \includegraphics[width=2in]{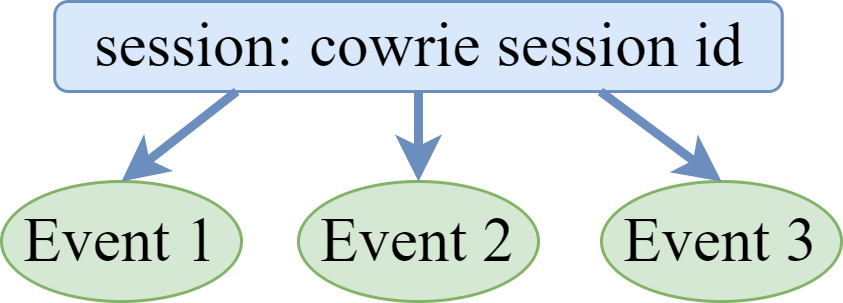}
    \centering
    \caption{A TAHOE \texttt{session} connects related \texttt{events}.}
    \label{fig:session}
\end{figure}

A TAHOE \texttt{session} is a special document that connects related events. Note that the event itself contains no information about any other event. However, a \texttt{session} contains references to all events that were generated during one Cowrie login session. Fig. \ref{fig:session} highlights how a TAHOE \texttt{session} groups related command \texttt{events}. Here, \texttt{events} $1, 2$ and $3$ were all recorded during one login session of an attacker. TAHOE maintains this relationship by connecting them to the same \texttt{session} node.

Also, Fig. \ref{fig:event} shows that a TAHOE \texttt{event} has a field called timestamp that stores the exact time when the command was input into the shell. Using these information we can sort the \texttt{events} in a \texttt{session} by their timestamps to get a sequence of commands input by an attacker. Table \ref{tbl:top5} shows the $5$ most popular command sequences along with their frequency in our dataset.

\section{Experimental Verification}

\begin{table*}[t]
    \caption{Top 5 most frequent command sequence of length $6$.}
    \label{tbl:top5}
    \centering
    \begin{tabular}{|c|c|}
        \hline
        Command Sequence & Frquency \\ \hline
        \texttt{shell - system - enable - /var/run/.ptmx - /etc/.ptmx - cp} & 654 \\ \hline
        \texttt{/usr/.ptmx - /boot/.ptmx - while - cp - chmod - cat'} & 653 \\ \hline
        \texttt{\%{input[0]} - /var/.ptmx - shell - system - enable - /var/run/.ptmx} & 656 \\ \hline
        \texttt{/dev/.ptmx - /dev/shm/.ptmx - rm - while - \%{input[0]} - /bin/.ptmx} & 662 \\ \hline
        \texttt{/var/tmp/.ptmx - /boot/.ptmx - chmod - /tmp/.ptmx - >/.ptmx - /usr/.ptmx} & 649 \\ \hline
    \end{tabular}
\end{table*}

To verify that our system works, we have done predictive analysis on command chains. Specifically, we have used Levenshtein distance \cite{yujian2007normalized} to predict the next command input by a bot. The high prediction accuracy of our model, as shown in subsection \ref{sec:res} shows that botnet behavior is predictable and our system can be used to analyze that.


\subsection{Training a Model}

As explained in subsection \ref{ss:seq} we form chains out of commands input into the Cowrie shell. An example of a command chain is \texttt{shell - /mnt/.ptmx - system - chmod - /boot/.ptmx - cat}. Here, the attacker inputs $6$ commands, in this particular order, before logging out. Let's replace these commands with integers in this example. Then the chain becomes \texttt{1-2-3-4-5-6}.

Assume, we have $N$ unique command chains in our dataset for $M$ sessions. We further divide these chains into sub-chains of lengths $3$ to $11$. For example, from the chain \texttt{1-2-3-4-5-6} we can get $3$ sub-chains of length $4$ -- \texttt{1-2-3-4, 2-3-4-5, 3-4-5-6} and $2$ sub-chains of length $5$ -- \texttt{1-2-3-4-5, 2-3-4-5-6}.

We have split the chains into sub-chains for two reasons -- (1) to obtain more data out of our dataset, (2) because our initial observation revealed that sub-chains of lengths $3$ to $11$ are common in many command chains in our dataset. However, this is purely a design choice and can be easily modified later.

\begin{figure}[!ht]
    \includegraphics[width=\linewidth]{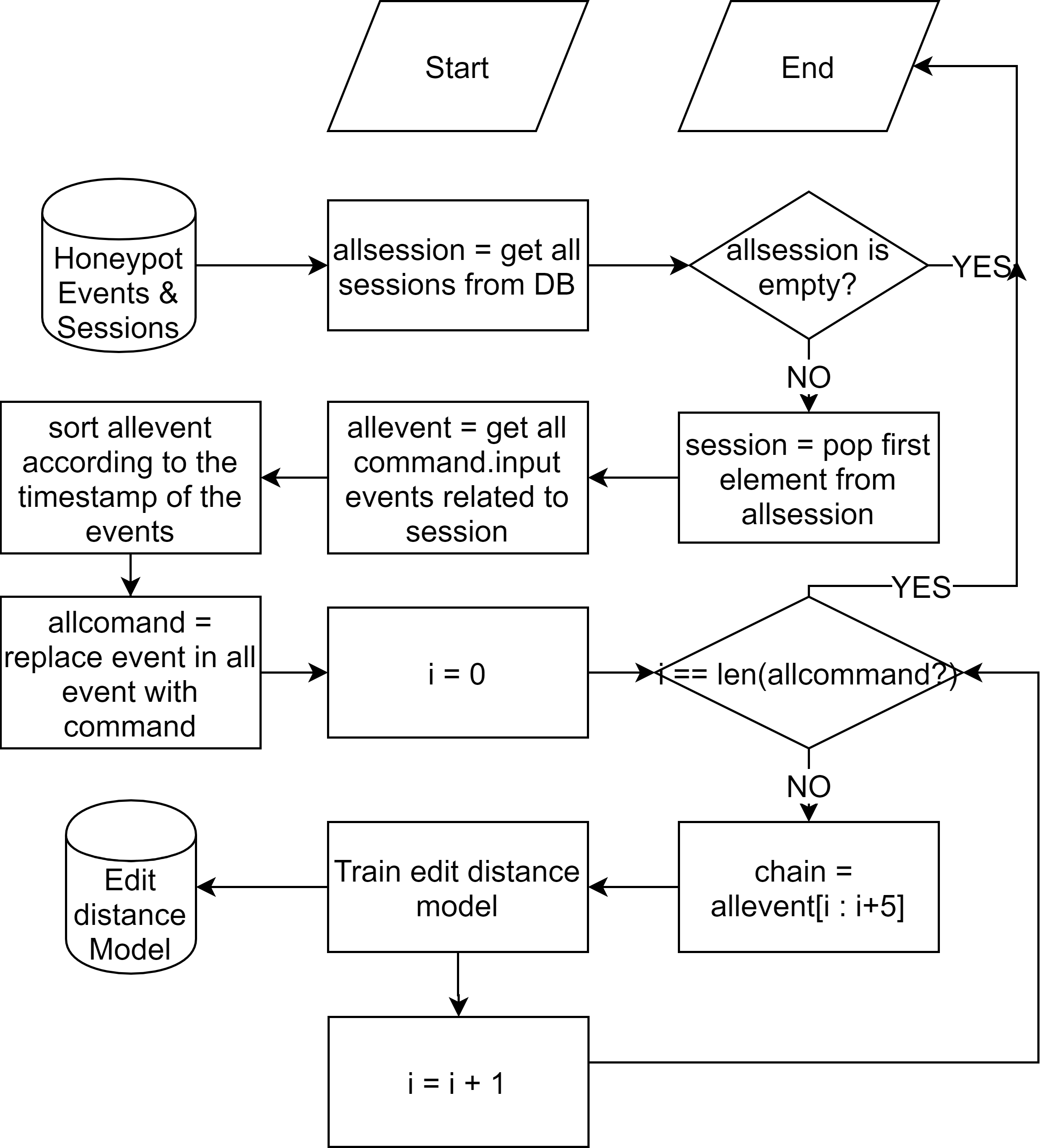}
    \centering
    \caption{Training methodology.}
    \label{fig:train}
\end{figure}

We have then created a model out of these command-chains by storing them in a nested hash table. The keys are the sub-chains except the last command. The corresponding value is a second dictionary. The keys of the second dictionary are the final command and values are the frequency of the entire sequence. Fig. \ref{fig:train} shows the training methodology for sub-chains of length $5$. This can easily be extended for other sub-chains of lengths $3$ to $11$. The methodology is explained below:

\begin{enumerate}
    \item Fetch all \texttt{sessions} from the database.
    \item For each \texttt{session}, fetch all of its related \texttt{events}.
    \item Select only \texttt{events} with sub\_type=`shell\_command'.
    \item Store the \texttt{events} in an array.
    \item Sort the array by the timestamp of the \texttt{events}.
    \item Assign an unique integer to each command. This array contains a sub-chain of a certain length ($5$ in this case).
    \item Feed the array into the edit distance training model. The model stores each sub-chain, the next command input by the attacker and the frequency of this sub-chain in our dataset.
\end{enumerate}

After doing the above operations, we end up with a number of arrays, each containing $5$ commands. One such array represents one sub-chain of commands input by the attacker. We also store their frequency or the number of times this sub-chain is seen in our dataset.

\subsection{Why Levenshtein distance?}

Levenshtein distance is a type of edit distance which quantifies how dissimilar two sequences (or strings) are. This is an industry standard technique to compare two sequences when one sequence can be obtained from the other by insertion, deletion or substitution of certain elements. There are other specialized types of edit distances like Longest common subsequence \cite{hunt1977fast} which only considers insertion and deletion but not substitution. Another example is Hamming distance \cite{hamming1950error} which only considers substitution, so it works for sequences of same length only. However in our case the two sequences can be totally different in content and length requiring the use of Levenshtein distance.

\subsection{Testing our Model}

We have split our total dataset into training and testing sets in a ratio of $80\%/20\%$. For testing, we have used normalized Levenshtein distance \cite{yujian2007normalized}. Levenshtein distance is a metric of similarity between two strings. We use this to detect the similarity between two chain of commands. It is a type of edit distance. A normalized Levensthein distance is calculated by dividing the total Levenshtein distance by the total length of both the sequences.

Now, to test a command chain, we perform the same operations as Fig. \ref{fig:train}. At the end of each loop we get an array of $5$ commands. The first $4$ commands form a sub-chain whereas the $5^{th}$ command is the actual next command. We then calculate the edit distance of the sub-chain with all the sub-chains in our model. We select the sub-chain with the least edit distance as the match.  If there are multiple matches we consider the one that has the highest frequency. A hash table lookup of the matched sub-chain gives us the predicted next command from our model. We can then test the accuracy of our system by comparing our predicted next command with the actual next command.

\section{Result}\label{sec:res}

\begin{figure}[!ht]
    \includegraphics[width=\linewidth]{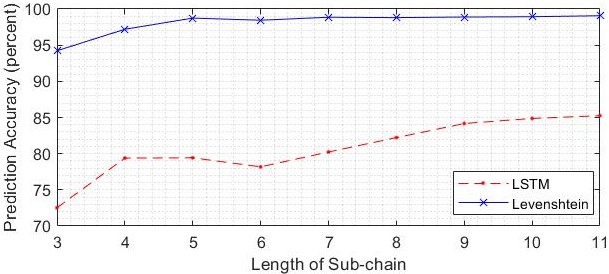}
    \centering
    \caption{Prediction Accuracy.}
    \label{fig:res}
\end{figure}

\begin{figure}[!ht]
    \includegraphics[width=\linewidth]{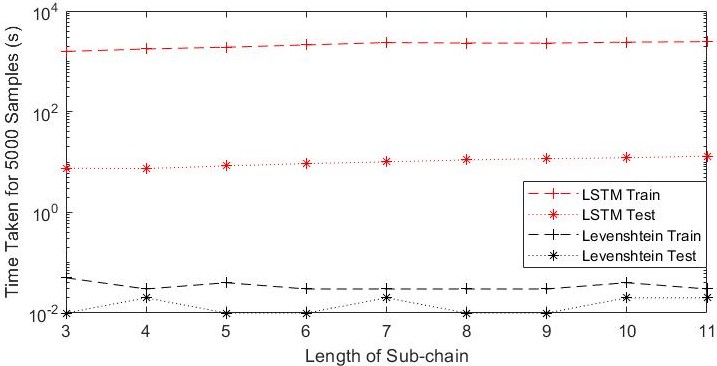}
    \centering
    \caption{Time Complexity of LSTM vs Levenshtein}
    \label{fig:complexity_5k}
\end{figure}

\begin{figure}[!ht]
    \includegraphics[width=\linewidth]{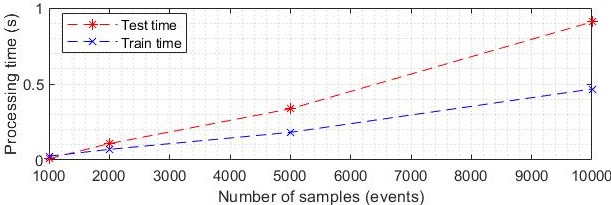}
    \centering
    \caption{Complexity of the system.}
    \label{fig:res_complexity}
\end{figure}

\begin{table*}
    \caption{Prediction Accuracy and Time Taken for $5000$ samples.}
    \label{tbl:result}
    \centering
    \begin{tabularx}{\linewidth}{|X|X|X|X|X|X|X|}
        \hline
        Length of Sub-chain &
            LSTM Accuracy (\%) &
            LSTM Train Time (s) &
            LSTM Test Time (s) &
            Levenshtein Accuracy (\%) &
            Levenshtein Train Time (s) &
            Levenshtein Test Time (s) \\

        \hline

        $3$ & $72.52$ & $1570.88$ & $7.53$ & $94.23$ & $0.05$ & $0.01$ \\ \hline
        $4$ & $79.39$ & $1758.75$ & $7.40$ & $97.20$ & $0.03$ & $0.02$ \\ \hline
        $5$ & $79.43$ & $1908.89$ & $8.37$ & $98.74$ & $0.04$ & $0.01$ \\ \hline
        $6$ & $78.18$ & $2117.82$ & $9.30$ & $98.45$ & $0.03$ & $0.01$ \\ \hline
        $7$ & $80.23$ & $2360.41$ & $10.04$ & $98.86$ & $0.03$ & $0.02$ \\ \hline
        $8$ & $82.22$ & $2308.45$ & $10.97$ & $98.81$ & $0.03$ & $0.01$ \\ \hline
        $9$ & $84.17$ & $2297.39$ & $11.68$ & $98.89$ & $0.03$ & $0.01$ \\ \hline
        $10$ & $84.87$ & $2408.08$ & $12.09$ & $98.93$ & $0.04$ & $0.02$ \\ \hline
        $11$ & $85.25$ & $2453.72$ & $13.07$ & $99.07$ & $0.03$ & $0.02$ \\ \hline
    \end{tabularx}
\end{table*}

Table \ref{tbl:result} shows the accuracy of the predictions of our training model. This same result is plotted in Fig. \ref{fig:res}. The result shows that, even after the attacker inputs only $3$ commands into the system, our methodology can predict with great certainty what the next command input by the attacker will be. This means that predictive analysis of attacker behavior works in this dataset. As the attacker keeps inputting more commands into the system, then length of the sequence increases and so does the prediction accuracy of our system.

Table \ref{tbl:result} and Fig. \ref{fig:res} also compares the accuracy of our model with Long Short Term Memory (LSTM). As seen our model (Levenshtein distance) has much better accuracy than LSTM. Furthermore, Table \ref{tbl:result} shows the superior performance of our methodology compared to LSTM. The training time of LSTM is much slower than our model and is not practical for real systems.

Fig. \ref{fig:complexity_5k} compares the time complexity of LSTM with that of Levenshtein distance. It shows that edit distance is much faster than LSTM during both training and testing phase.

Fig. \ref{fig:res_complexity} shows the time complexity of our training and testing procedures. Note that, the x-axis of Fig. \ref{fig:res_complexity} is sample size not sub-chain length. The result shows that the process is linear and thus can be scaled horizontally. This is important for a realtime system that collects data from thousands of hosts. As the whole process from data collection to behavioral analysis scales linearly, we can use clustering or map-reduce to support virtually infinite hosts.

\section{Conclusion and Future Work}
In this work we have outlined a robust framework for automated analysis of botnet behavior. To do this, We have used a novel approach of analyzing chain of commands input by attackers into the shell. We have also incorporated a sequential analysis to verify that our approach can correctly predict attacker behavior. A prediction accuracy of $94-99\%$ proves the validity of our approach.

In future we will look into real-time classification by using both benign and malicious datasets. We would then try to make the methodology more versatile by utilizing dataset generated by other malware not only Mirai. We will also cross validate our procedure with a newer dataset collected over a different time span.

\bibliographystyle{IEEETran}
\bibliography{refs}

\end{document}